\documentclass{aastex}          
\usepackage{spr-astr-addons}    
\usepackage{url}

\newcommand{\emaila}{yoelsy@uclv.edu.cu}

\begin{document}
%
\title{Quinstant Dark Energy Predictions for Structure Formation}

\shorttitle{Quinstant Dark Energy Predictions for Structure Formation}
\shortauthors{<Yoelsy Leyva et al.>}

\author{Yoelsy Leyva Nodal \altaffilmark{1}}
\and
\author{Rolando Cardenas \altaffilmark{1}}
\affil{Departamento de Fisica, Universidad Central de las
Villas, Santa Clara, CP, 54830, Villa Clara, Cuba}
\and
\author{V.F. Cardone \altaffilmark{2}}
\affil{Dipartimento de Fisica ¨E.R.Caianiello¨, Universita di
Salerno and INFN, Sezione di Napoli, Gruppo Collegato si Salerno,
Via S. Allende, 84081 - Baronissi(Salerno), Italy}
\email{\emaila}


\begin{abstract}
We explore the predictions of a class of dark energy models, quinstant dark energy, concerning the structure formation in the Universe, both in the linear and non-linear regimes. \textit{Quinstant}  dark energy is considered to be formed by \textit{quin}tessence and a negative cosmological con\textit{stant}. We conclude that these models give good predictions for structure formation in the linear regime, but fail to do so in the non-linear one, for redshifts larger than one.

\end{abstract}


\section{Introduction}
In a former publication (\cite{Cardone2008}) we explored the potential degeneracy existing between some dark energy models and $f(R)$ modified gravity theories. This means that in principle we might reproduce the dynamics of the expansion of the Universe with a dark energy model, and then obtain an equivalent $f(R)$ model with the same dynamics. In that paper we worked with a class of dark energy models which could be dubbed \textit{quinstant} dark energy, as this ingredient is considered to be a composite of a \textit{quin}tessence scalar field and a (negative) cosmological con\textit{stant}. We refer the interested reader to (\cite{Cardenas2002a}, \cite{Cardenas2002b}, \cite{Cardenas2003}, \cite{Cardone2008}) for details on quinstant dark energy.

In that work we also wrote on the potential of structure formation to break away the above mentioned degeneracy: the same expansion history might not imply the same formation history. Therefore, in this paper we give a step forwards  concerning the predictions of quinstant dark energy for large scale structure formation, leaving for the future the exploration of the (potentially more difficult) predictions of $f(R)$ modified gravity theories.

\section{Some characteristics of quinstant dark energy}

We investigate spatially flat, homogeneous and isotropic
cosmological models (\cite{Cardenas2003, Cardone2008}) filled with three non-interacting components:
pressureless matter(dust), a scalar field $\phi$ and a negative
cosmological constant $\Lambda$. We first consider the potential
introduced by \cite{Rubano2002, Cardenas2002}:
\begin{equation}
    V(\phi)\propto exp \{ -\sqrt{\frac{3}{2}}\phi\}
\end{equation}

For this potential the following substitutions inspired in the
Noether symmetric approach allow analytical solutions:
\begin{eqnarray}
  a^{3} &=& u v \\
  \phi &=& -\sqrt{\frac{2}{3}}ln(\frac{u}{v})
\end{eqnarray}

where $a$ is the scale factor, $u_{1}$ and $u_{2}$ are functions of
time. This makes it possible to integrate the Friedmann equations exactly.
Setting at present ($\tau=0$): $a(0)=1$, $\dot{a}(0)=1$ and $H(0)=1$
we obtained the evolution of the scale factor as a function of time:
\begin{equation}
   a=\left[\frac{cos^{2}\omega \tau}{2\omega \sqrt{-\Omega_{\Lambda}}}(a_{1}+
   b_{1}tan\; \omega\tau)(a_{2}+b_{2}tan \;\omega\tau)         \right]
\end{equation}

with:
\begin{eqnarray}
  a_{1} &=& 1 \\
  b_{1} &=& u_{1}/u_{2} \\
  a_{2} &=& w(2\sqrt{-\Omega_{\Lambda}}-u_{1}u_{2}\tau )\\
  b_{2} &=& 2\sqrt{-\Omega_{\Lambda}}(3-u_{1}\omega/u_{2})+u_{1}u_{2}+u_{2}^{2}\omega\tau\\
  w     &=& 2\sqrt{-\Omega_{\Lambda}}/2
\end{eqnarray}

\section{Parameterisation of the equation of state}
Because of the rapid evolution of the equation of state in our
solutions, the conventional parameterisations (\cite{Huterer:1998qv, Huterer:2000mj, Weller:2000pf,
Chevallier:2000qy, Linder:2004ng, Linder:2002et}) do not work properly.  In order to
parameterise it we use a 4-parameters equation of state proposed by
\cite{Linder:2005ne}. This is similar to the \textit{Kink} aproach
proposed by \cite{Bassett:2004wz, Corasaniti2003, Corasaniti2003a}.

\begin{equation}\label{1}
    w(a)=w_{0}+ \frac{w_{m}-w_{0}}{1+(\frac{a}{a_{T}})^{1/\tau}}
\end{equation}

Here $a_{T}$ is the value of the scale factor at a transition point
between $w=w_{m}$, the value at the matter dominated era, and
$w=w_{0}$, the value today. The rapidity of the evolution is
measured by $\tau$

\begin{figure}[h]
\includegraphics[width=1.0\linewidth]{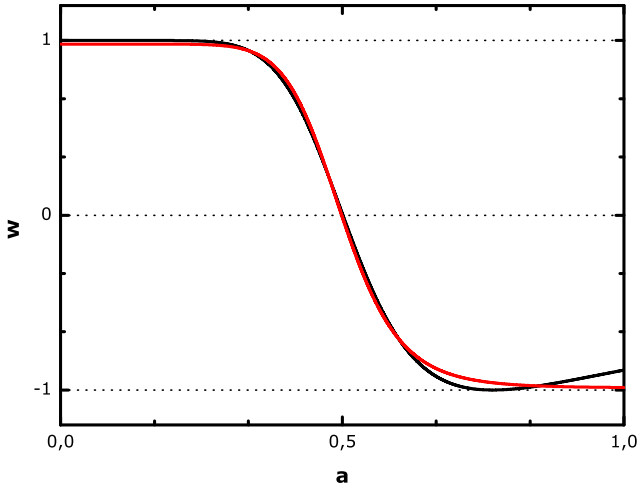}
\caption{Behaviour of the equation of state (here called $w$) versus
the scale factor. The red line corresponds to the $4-$parameters
approach. The black dotted lines are the limit cases, from top to bottom: \textit{a)} $w=1$ stiff matter,
\textit{b)} $w=0$ dust, \textit{c)} $w=-1$ cosmological constant.} \label{fig:trackex}
\end{figure}

Fig. \ref{fig:trackex} shows the parameterisation. The best fit is
obtained with the following values of the parameters:
\begin{eqnarray}
  w_{0} &=& -0.98794 \pm 0.00064\\
  w_{m} &=&  0.97944 \pm 0.00097\\
  a_{T} &=&  0.49939 \pm 0.00012\\
  \tau  &=&  0.10254\pm 0.00021
\end{eqnarray}

This approach allows integration over $w$ to be done analytically:
\begin{equation}\label{f(a)}
    f(a)=exp\left[3\int_{a}^{1}\left( \frac{1+3w(u)}{u}\right)du \right]
\end{equation}

and so the Hubble parameter can be written explicitly:
\begin{eqnarray}\label{hubble}
  (\frac{H}{H_{0}})^{2} &=& \Omega_{m0}a^{-3}+(1-\Omega_{m0})a^{-3(1+w_{m})} \nonumber\\
 & &\left(1+a_{T}^{-1/\tau}\right)^{-3\tau \Delta w} \left(1+(\frac{a}{a_{T}})^{1/\tau}\right)^{3\tau \Delta w}
 \label{H}
\end{eqnarray}

\section{Linear growth of fluctuations}
The dynamics of these models has been successfully probed using the
standard test of the acoustic peak, the dimensionless coordinate
distance $y(a)$ and others. In this work we must study the role of dark
energy on the evolution of density perturbations.

\begin{figure}[h]
\includegraphics[width=1.0\linewidth]{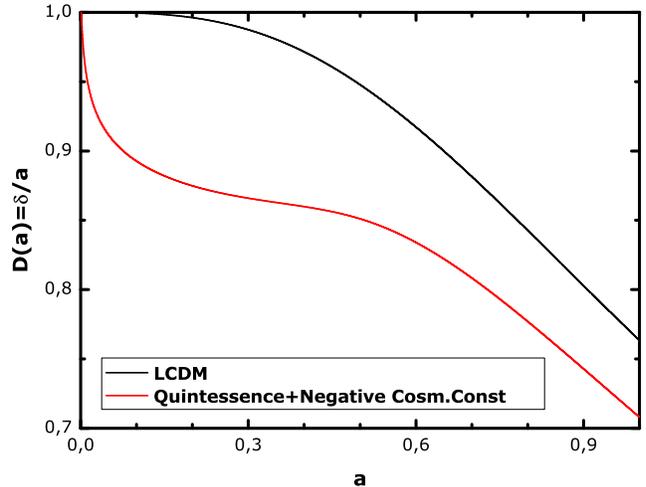}
\caption{The growth history($D(a)=\delta/a$) is shown for the
composite dark energy model(red line). The black line corresponds to
the $\Lambda$CDM model} \label{Growth}
\end{figure}

The perturbation equation reads as:
\begin{equation}\label{7}
    \ddot{\delta}+2H\dot{\delta}-4\pi G\rho_{M}\delta=0
\end{equation}

It is convenient to study the growth evolution in terms of the scale
factor $a$ or the characteristic scale $ln\; a$
rather than time $t$. Because in the matter domination
epoch the solution is $\delta(a) \sim a$, it is useful in studying the influence of
dark energy to divide out this behavior and shift to the growth
variable $D(a)=\frac{\delta}{a}$. So we can rewrite equation
(\ref{7}) as:
\begin{eqnarray}\label{final}
&& D''+D'\left[\frac{5}{a}+\frac{(ln\;E^2)'}{2}\right]\nonumber\\
&&+\frac{D}{a}\left[
\frac{3}{a}\left(1-\frac{\Omega_{M}}{2E^{2}a^{3}} \right)+
\frac{(ln\;E^2)'}{2} \right]=0
\end{eqnarray}

where the prime denotes derivation with respect to the scale factor
$a$ and $ E^{2}\equiv(\frac{H}{H_{0}})^{2}$. This equation is
solved numerically using the boundary conditions $D(a_{LSS})=1$ and
$D'(a_{LSS})=0$. Here $a_{LSS}=\frac{1}{1+z_{LSS}}$ denotes the
scale factor evaluated at the redshift of the last scattering
surface. In our calculation we used the approximated value of
$z_{LSS}\approx 1089$

\begin{figure}[h]
\includegraphics[width=1.0\linewidth]{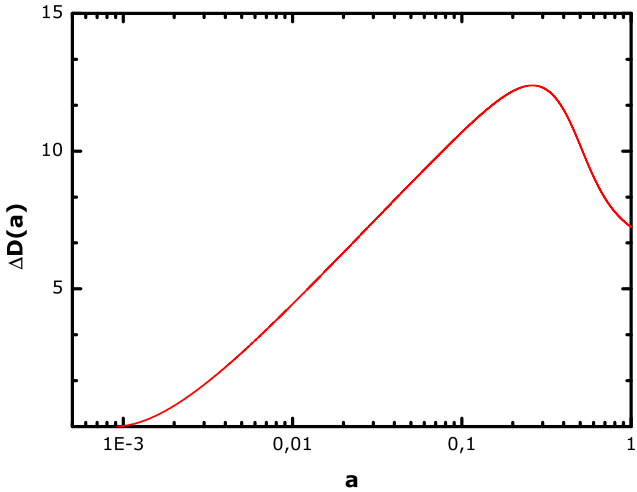}
\caption{Percent deviation of the growth variable $D$ from the $\Lambda$CDM predictions}
\label{Delta}
\end{figure}

The growth history is shown in Fig. \ref{Growth}, more information
is given by Fig. \ref{Delta}, which represents the percentage
deviation ($\Delta D(a)=1-D(a)/D_{\Lambda}(a)$) of the growth factor
for the composite model of dark energy with respect to that for the
concordance $\Lambda CDM$ one as a function of the scale factor.
Current deviations are less than $8\%$.

In the linear perturbation theory the growth index can be
measured by the peculiar velocity field \textbf{v}. This quantity
is defined as:
\begin{equation}\label{growth index}
    f=\frac{d\ln \delta}{d\ln a}
\end{equation}

Nevertheless, it is better to solve a direct equation for $f$, to avoid
propagating numerical errors from the calculation of $\delta$. For
this purpose, one should simply use the definition of $f$ across
(\ref{growth index}) and Eq.(\ref{final}) to get the evolution
equation of the growth index versus scale factor:
\begin{equation}\label{growth index}
    f'+\frac{f^{2}}{a}+ \left[\frac{2}{a}+\frac{(ln\;E^{2})'}{2}
    \right]f-\frac{3\Omega_{M}}{2E^{2}a^{4}}=0
\end{equation}

\begin{figure}[h]
\includegraphics[width=1.0\linewidth]{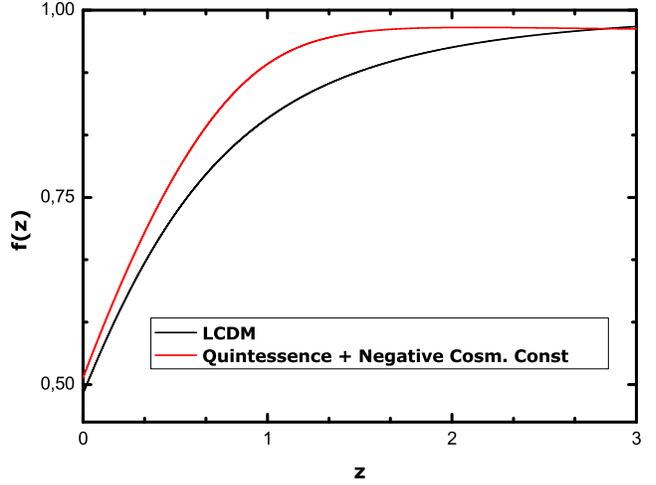}
\caption{The growth index as a function of the scale factor for the
composite dark energy model(red line). The black line corresponds to
the $\Lambda$CDM model }\label{f}
\end{figure}

\begin{figure}[h]
\includegraphics[width=1.0\linewidth]{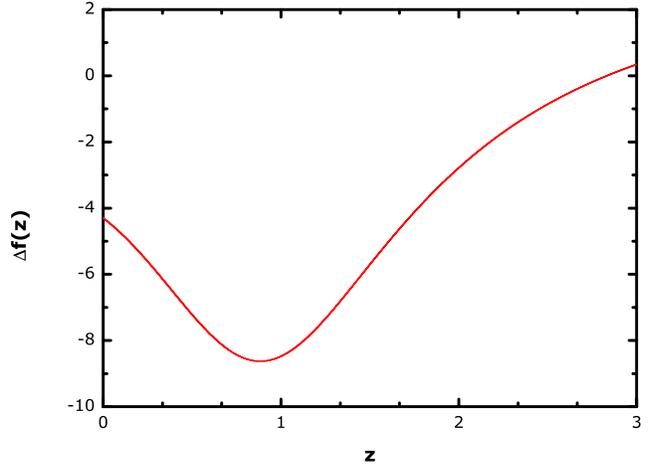}
\caption{Percent deviation of the growth index $f$ from the $\Lambda$CDM predictions}\label{DeltaF}
\end{figure}

This equation is solved numerically using the initial condition
$f(a_{LSS})=1$ and Eq.(\ref{H}). The result is shown in
Fig.~\ref{f}. To analyse the behavior of $f$ predicted by the
concordance scenario and the composite dark energy model we compute
the percentage deviation between their predictions as $\Delta
f(z)=1-f(z)/f_{\Lambda}(z)$. This is shown in Fig.\ref{DeltaF}. It is
worth noting that $\Delta$f is always negative, i.e., the growth
index of the composite dark energy model is larger than the
$\Lambda$CDM. Despite this result the predicted value of the growth
index at the survey effective depth $z=0.15$ is in very good
agreement with the estimated value reported in \cite{Sen2006}

\section{Non linear evolution of a spherical fluctuations}
Another powerful tool to understanding how a small spherical patch of homogeneous overdensity forms a bound system via gravitational instability is the spherical collapse model(see \cite{Basilakos:2003bi},  \cite{Mota:2004pa},\cite{Horellou2005a}, \cite{Maor2005}, \cite{Wang2006}). The equation of motion of a spherical shell in the presence of dark energy is
\begin{equation}\label{over}
    \frac{\ddot{r}}{r}=-\frac{4\pi G}{3}(\rho_{cluster}+ \rho_{X,eff})
\end{equation}
where $\rho_{cluster}$ is the time varying density inside the forming cluster and
\begin{equation}\label{rho}
    \rho_{X,eff}=\rho_{X}+3p_{X}=(3w(r)+1)\rho_{X}
\end{equation}
is the effective energy density of the X$-$component. For a cosmological constant case $\rho_{X,eff}$ is constant. Note that for $w(r)\neq -1$ and $w(r)\neq -1/3$ the density of the X$-$component inside the overdensity patch depends on the evolution of the background. If the initial overdensity exceeds a critical value, the overdensity region will break away from the general expansion and go through an expansion up to a maximum radius (turnaround point), then it will begin to collapse until the sphere virializes forming a bound system.
\subsection{Rescaled equations and evolution until turnaround}

Let us perform the following transformations:
\begin{equation}\label{rescaled1}
    x=\frac{a}{a_{ta}}
\end{equation}
and
\begin{equation}\label{rescaled2}
    y=\frac{r}{r_{ta}}
\end{equation}
where $a_{ta}$ is the scale factor of the background and $r_{ta}$ is the radius of the overdensity when the perturbation reaches the turnaround. In a flat cosmology ($k=0$), using equations (\ref{hubble},\ref{over}), the evolution of above quantities is given by:
\begin{equation}\label{x_evol}
    \dot{x}=H_{ta}\Omega_{ta}^{1/2}[\Omega(x)x]^{-1/2}
\end{equation}
\begin{equation}\label{y_evol}
    \ddot{y}=-\frac{H_{ta}\Omega_{ta}^{1/2}}{2}\left[ \frac{\zeta}{y^{2}}+\frac{\Omega_{X,ta}}{\Omega_{ta}}y[1+3w(x)]f(x)
      \right]
\end{equation}
where we have considered that the spherical overdensities evolve in the presence of homogeneous
dark energy (generally dark energy does not cluster on scales smaller than $100$ $Mpc$, see \cite{Dave2002a}) and also the mass of the forming cluster is conserved:
\begin{equation}\label{cons}
    \rho_{cluster}r^{3}=\rho_{cluster,ta}r_{ta}^{3}
\end{equation}
and
\begin{equation}\label{sobred}
    \zeta=\left(\frac{\rho_{cluster}}{\rho_{backg}} \right)_{z_{ta}}
\end{equation}
is the overdensity of the forming cluster at turnaround. Its value can be calculated by integrating the above differential equations(\ref{x_evol},\ref{y_evol}) using at the same time the boundary conditions $(dy/dx)_{x=1}=0$ and $y_{x=0}=0$.

\begin{figure}[h]
\includegraphics[width=1\linewidth]{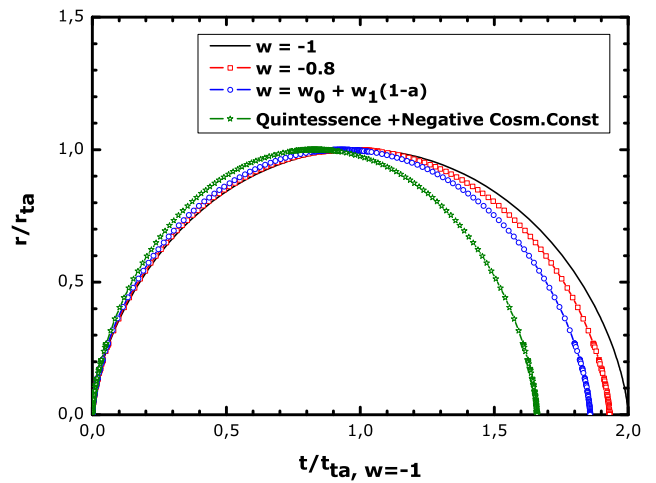}
\caption{Evolution of a perturbation collapsing at the present time ($z_{coll}=0$) in different cosmologies. The values $w_{0}=-0.75$ and $w_{1}=0.4$ were used in the third dark energy parametrization. The x$-$ axis shows the time normalised to the turnaround time for the $\Lambda$CDM model.}\label{Evol_Pert}
\end{figure}

\begin{figure}[h]
\includegraphics[width=1.0\linewidth]{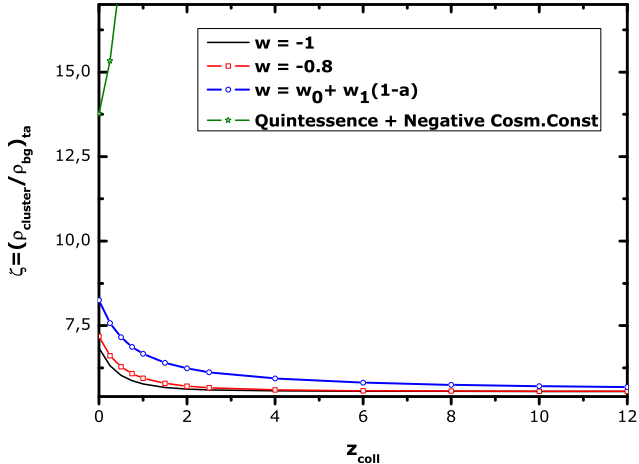}
\caption{Evolution of a perturbation collapsing at the present time ($z_{coll}=0$) in different cosmologies. The composite dark energy model, upper left side of this figure, has an unusual behavior}\label{Zeta}
\end{figure}

The variations of a perturbation of a collapsing now in different cosmologies are compared in Fig. \ref{Evol_Pert}. The composite dark energy model reaches turnaround and collapse earlier than the other models, this behavior is because of the effect of the negative cosmological constant which contributes with a positive pressure in the whole evolution. The evolution of the overdensity $\zeta$ with collapse redshift is shown in Fig. \ref{Zeta}. All quintessence models, including the $w=w_{0}+w_{1}(1-a)$ parameterisation, show that perturbations collapsing at certain redshift are denser at turnaround, relative to the background, than in the $\Lambda$CDM model. These models tend at high redshift to the fiducial value of $5.6$ for the Einstein-de Sitter universe. Nevertheless, the composite dark energy model shows an unusual evolution because its relative overdensities grow while the redshift tends towards a higher values.

\begin{figure}[h]
\includegraphics[width=1.0\linewidth]{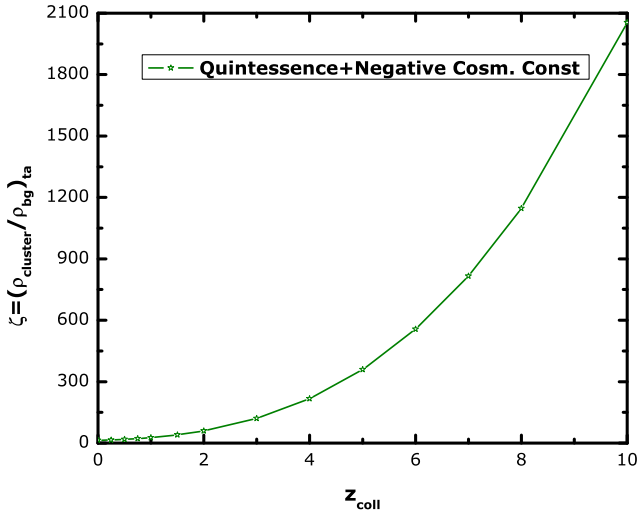}
\caption{Evolution of a perturbation collapsing at the present time ($z_{coll}=0$) for the quinstant dark energy model. The overdensity grows indefinitely as $z_{coll}$ tends to high redshift.}\label{Compo}
\end{figure}

This worrying behavior is more evident in Fig. \ref{Compo}. This is a first important departure of quinstant dark energy from the mainstream interpretation of astrophysical observations.

\subsection{Density contrast at virilization}
In order to obtain more information about the predictions of the composite dark energy model from the growth history point of view, we need to compute the density contrast of the virilised cluster as a function of the collapse redshift:
\begin{equation}\label{contrast}
    \Delta_{vir}=\frac{\rho_{vir}}{\rho_{bg,vir}}=\eta^{-3}\zeta \left( \frac{1+z_{ta}}{1+z_{coll}}\right)^{3}
\end{equation}
where $\eta=\rho_{vir}/\rho_{ta}$ is the ratio of density of the virialised sphere to the density at turnaround. This parameter is very sensitive to the way you consider the virialisation process. The usual method to find $\eta$ is combining the virial theorem\footnote{both components of the quinstant dark energy model have their potential energies proportional to $R^{2}$} and conservation of energy to obtain a relation between the potential energies of the collapsing sphere at turnaround and at virialisation time.
In this process the gravitational potential energy of the spherical dark matter
overdensity will be modified by a new term due to the gravitational effects of dark energy
on dark matter(see \cite{Horellou2005a} and references therein).

When dark energy is dynamical, i.e. quintessence, the total energy of the bound system is not conserved and will contribute as a non-conservative force to the dark matter particles (see \cite{Maor2005, Zeng:2005ib}). One way to avoid this issue in the context of an homogeneous dark energy component was proposed in \cite{Wang2006}.This approximation considered that dark matter particles can reach a quasi-equilibrium state in which virial theorem holds instantaneously because during the matter domination era the effect of dark energy is still small.

\begin{figure}[h]
\includegraphics[width=1\linewidth]{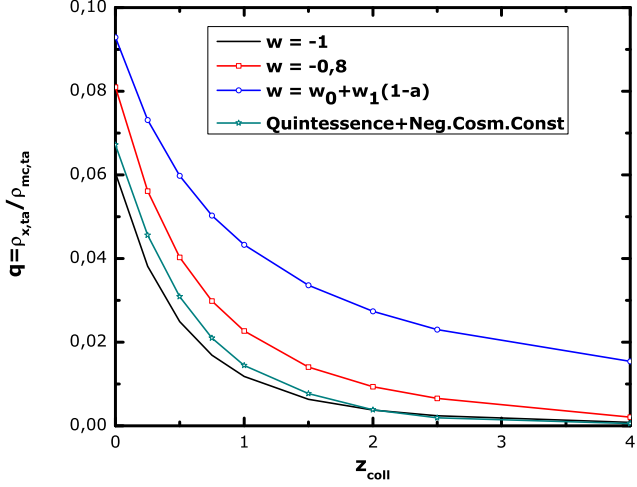}
\caption{Evolution of the parameter $q$ respect to the virialisation redshift $z_{vir}$.}\label{q vs.zcoll}
\end{figure}

Thus assuming dark matter has reached this quasi-equilibrium state, its total energy can be
computed by the virial theorem :
\begin{equation}\label{P.wang}
4q(1+3w_{x,vir})\eta^{3}-2(1+(1+3w_{x,ta})q)x+1=0
\end{equation}
where $q=\rho_{x,ta}/\rho_{mc,ta}$ characterises the strength of dark energy at turnaround.
Fig. \ref{q vs.zcoll} shows the evolution of $q$ on $z_{coll}$ for the composite dark energy model. The effect of energy nonconservation is not so worrying because $q$ will always be quite small

\begin{figure}[h]
\includegraphics[width=1\linewidth]{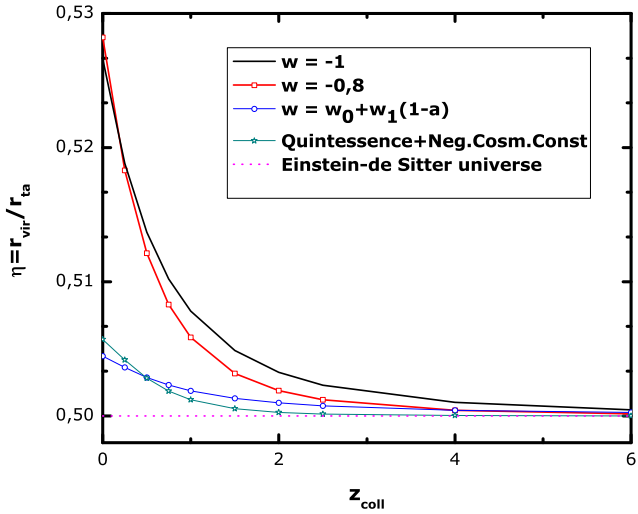}
\caption{Ratio of the virialization radius to the turnaround radius $\eta=r_{vir}/r_{ta}$ as a function of the $z_{vir}$.}\label{x vs zcoll.eps}
\end{figure}

The variations of the ratio of the virialisation radius to the turnaround radius is shown in Fig. \ref{x vs zcoll.eps}. The shape of the composite dark energy evolution is over the standard Einstein-de Sitter universe(pink dotted line), the value of $\eta > 0.5$ is due to the fact that the effective repulsive force of dark energy on dark mater particles allows them to reach the quasi equilibrium with a larger radius. This behavior is less marked in the composite dark energy model than the rest of the models because the negative cosmological constant contributes to the evolution with a positive pressure in contrast to the negative one of the quintessence.

\begin{figure}[h]
\includegraphics[width=1\linewidth]{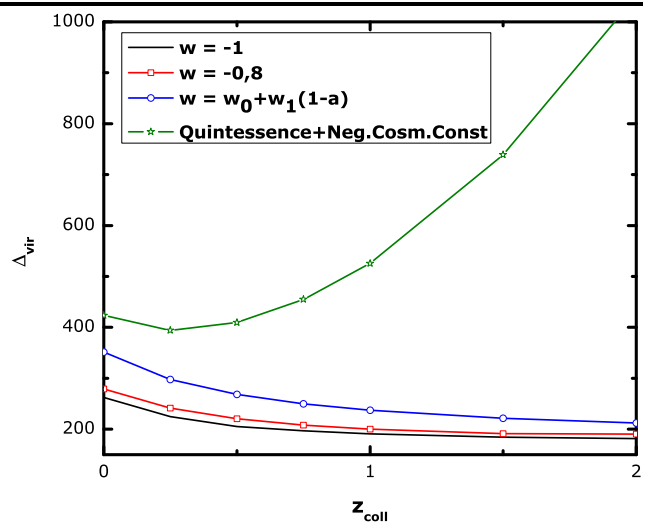}
\caption{Overdensity $\Delta_{vir}$ as a function of the collapse redshift for some cosmological models. The composite dark energy model shows an unusual evolution.}\label{overdesity}
\end{figure}

Finally, the density contrast at virialisation is shown in Fig. \ref{overdesity}. All the models have overdensities that decrease with the value of the $z_{coll}$ but the composite dark energy has the opposite behavior, so this result is consistent with the previous one shown in Fig. \ref{Zeta} and Fig. \ref{Compo}.

\section{Cluster Number Counts}
Let us now study the cluster abundance in the Universe using the Press-
Schechter formalism\footnote{it is possible to use more realistic mass function(see \cite{Sheth:1999mn} and \cite{Jenkins:2000bv}) but the aim of this section is to draw a comparison between the general predictions for the cluster number counts of the composite dark energy model and $\Lambda$CDM }. Following this approach the comoving number density of collapsed halos with masses in the range $M$ and $M + dM$ at a given redshift z can be calculated using the standard equation:
\begin{equation}\label{P-S}
\frac{dn}{dM}=-\sqrt{\frac{2}{\pi}}\frac{\rho_{m0}}{M}\frac{\delta_{c}}{\sigma(M,z)}\frac{d ln \sigma(M,z)}{dM}e^{-\frac{\delta^{2}(z)}{2\sigma^{2}(M,z)}}
\end{equation}
where $\rho_{m0}$ is the present average matter density of the Universe, $\delta_{c}(z)$ is the linear overdensity of a perturbation collapsing at redshift $z$ and $\sigma(M,z)=D(z)\sigma_{M}$ is the \textit{rms} density fluctuation in spheres of comoving radius R containing the mass M.

We have used the fit provided by \cite{Weinberg:2002rd} to $\delta_{c}(z)$ because this parameter shows a weak dependance on the cosmological constant.
\begin{equation}\label{kamio1}
    \delta_{c}(z)=\frac{3(12\pi)^{2/3}}{20}[1+\alpha Log_{10} \Omega_{m}(z)]
\end{equation}
\begin{equation}\label{kamio1}
    \alpha=0.353w^{2}+1.044w^{3}+1.128w^{2}+0.555w+0.131
\end{equation}

\begin{figure}[h]
\includegraphics[width=1\linewidth]{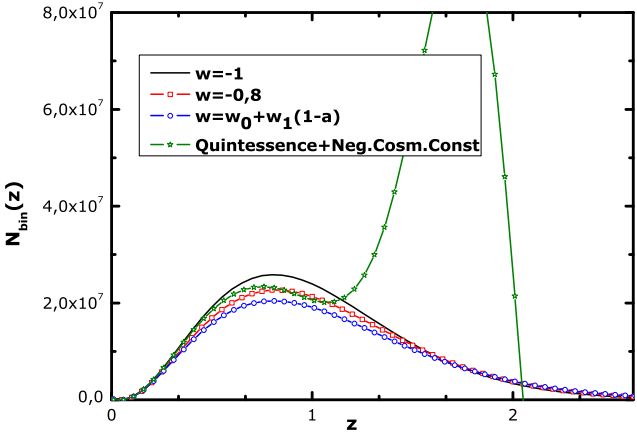}
\caption{Evolution of number counts in mass bins with redshift for objects with mass
within the range $10^{13} < M/(h^{-1}M_{\odot}) < 10^{14}$.}\label{cluster1}
\end{figure}

We use the fit provided in \cite{Viana:1995yv} to obtain the quantity
\begin{equation}\label{sigma8}
    \sigma_{M}=\sigma_{8}\left( \frac{M}{M_{8}}\right)^{-\frac{\gamma(M)}{3}}
\end{equation}
where $\sigma_{8}$ is the variance of the over-density field smoothed on a scale of size $R_{8}=8h^{-1}Mpc$, its actual value is constrained by the observations of $WMAP$ (\cite{Komatsu2009a}) and $M_{8}=6$ x $10^{14}\Omega_{m0}h^{-1}M_{\odot}$. The index $\gamma$ is a function of the mass scale and the shape parameter ($\Gamma=0.149$) of the matter power spectrum (see \cite{Viana:1995yv} for a detailed derivation of these quantities):
\begin{equation}\label{gamma}
    \gamma(M)=(0.3\Gamma + 0.2)\left[ 2.92 + \frac{1}{3}\log\left(\frac{M}{M_{8}}\right)\right]
\end{equation}

\begin{figure}[h]
\includegraphics[width=1\linewidth]{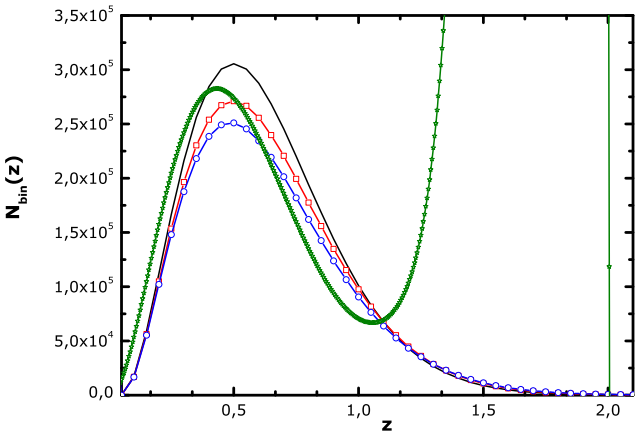}
\caption{Evolution of number counts in mass bins with redshift for objects with mass
within the range $10^{14} < M/(h^{-1}M_{\odot}) < 10^{15}$. Lines are the same as in Fig. \ref{cluster1}.}\label{cluster2}
\end{figure}

In order to compare the different models we use the normalisation equation (label \textit{mod} stands for model):
\begin{equation}\label{comp}
    \sigma_{8,mod}=\frac{\delta_{c,M}(z=0)}{\delta_{c,\\Lambda}(z=0)}\sigma_{8,\Lambda}
\end{equation}
The value of $\sigma_{8,\Lambda}=0.796$ is also constrained by WMAP (\cite{Komatsu2009a}) .
In this work the effect of dark energy on the number of dark matter halos is analyzed by computing the number of halos per unit of redshift, in a given mass bin
\begin{equation}\label{bin}
    N_{bin}=\frac{dN}{dz}\int_{4\pi}d\Omega\int_{M_{inf}}^{M_{sup}}\frac{dn}{dM}\frac{d^{2}V}{dzd\Omega}dM
\end{equation}
where the comoving volume is
\begin{equation}\label{comoving}
    \frac{d^{2}V}{dzd\Omega}=c\frac{d_{A}(z)^{2}(1+z)^{2}}{H(z)}
\end{equation}
and $d_{A}(z)$ is the angular diameter distance.

\begin{figure}[t]
\includegraphics[width=1\linewidth]{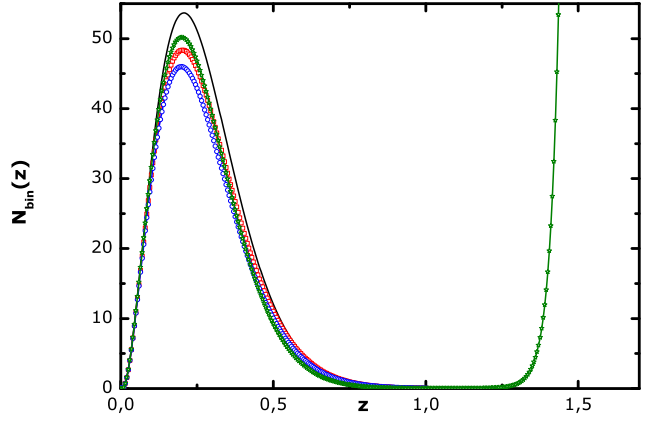}
\caption{Evolution of number counts in mass bins with redshift for objects with mass
within the range $10^{15} < M/(h^{-1}M_{\odot}) < 10^{16}$. Lines are the same as in Fig. \ref{cluster1}.}\label{cluster3}
\end{figure}

The number counts in mass bins, $N_{bin} = dN/dz$, obtained from (\ref{bin}), are shown
in Figures. \ref{cluster1}, \ref{cluster2}, \ref{cluster3}. All the models exhibit similar behavior:
the more massive structures are less abundant and form at later
times, as it should be in the hierarchical model of structure formation (see \cite{Liberato2006} for a further discussion in another dark energy parametrization).

Concerning the quinstant dark energy model, it is capable of reproducing the results of the other models in a satisfactory way backwards in time up to redshifts a bit larger than $z=1$ for the three range of mass values. Then, it shows abrupt peaks of structure formation, in a serious departure of the hierarchical model for large scale structure. This seems to be caused by the unusual equation of state of quinstant dark energy, which behaves as stiff matter for redshifts a bit larger than one. This would result in enhanced accretion of the forming structures, both because of gravitational and viscous forces.

\section{Conclusions}

Our opinion is that any dark energy model which shows a stiff-like equation of state in the past, during a long period of time, will predict abrupt peaks of structure formation, which  would be the result of enhanced accretion of the forming structures, both because of gravitational and viscous forces. This seems to be the case for all studied models of quinstant dark energy (\cite{Cardenas2002a}, \cite{Cardenas2002b}, \cite{Cardenas2003}, \cite{Cardone2008}). There are other cosmological models of composite dark energy having stiff-matter domination as an attractor in the past, but usually they would not be global attractors, but local. This is the case of several models of quintom dark energy.

Our general conclusions on structure formation predictions of quinstant dark energy (QDE) are:

\begin{itemize}
    \item QDE makes reasonable predictions for the formation of linear large scale structure of the Universe.
    \item It reproduces reasonably well the non-linear structures from today up to redshifts a bit larger than one
    \item   It fails to reproduce the perturbations in the non-linear regime for redshifts a bit larger than one.
    \end{itemize}

In the future we would study how the dynamically equivalent models of $f(R)$ modified gravity would behave concerning structure formation, and then we would have a better understanding on how these observations would crack the degeneracy dark energy-$f(R)$ modified gravity.

\section{Acknowledgments}
We acknowledge the Ministry of Higher Eduacation of Cuba for partial
financial support of this research, and Cathy Horellou for instructive comments and suggestions on this work.

\bibliographystyle{spr-mp-nameyear-cnd}
\bibliography{bib}

\end{document}